\documentclass[aps,pre,groupedaddress,amsmath,eqsecnum,floatfix]{revtex4}
\usepackage{graphicx} %(voor importeren van eps files)
\usepackage[latin1]{inputenc}
   \usepackage{bm}
\newcommand{\Label}[1]{\label{#1}}                  %% DO NOT show labels
\newcommand{\Bibitem}[1]{\bibitem{#1}}        %% DO NOT show labels
\newcommand{\be}{\begin{equation}}
\newcommand{\ee}{\end{equation}}
\newcommand{\ba}{\begin{eqnarray}}
\newcommand{\ea}{\end{eqnarray}}

\newcommand{\nn}{\nonumber\\}
\newcommand{\Ref}[1]{(\ref{#1})}
\newcommand{\dav}[1]{\langle \!\langle #1{\rangle \!\rangle}}  %type \av{A} to make <A>
\newcommand{\avo}[1]{\langle #1{\rangle}_o}

\newcommand{\sav}[1]{\langle #1{\rangle}^{(2)}}
\newcommand{\half}{\textstyle{\frac{1}{2}}}

\newcommand{\fifth}{\textstyle{\frac{1}{5}}}
\newcommand{\eighth}{\textstyle{\frac{1}{8}}}

\newcommand{\sig}{\sigma}
\newcommand{\bsig}{{\bf \sig}}
\newcommand{\hsig}{\hat{\sig}}
\newcommand{\hbsig}{\hat{\bsig}}

\newcommand{\bv}{{\bf v}}

\newcommand{\bg}{{\bf g}}
\newcommand{\br}{{\bf r}}
\newcommand{\bG}{{\bf G}}
\newcommand{\bR}{{\bf R}}

\newcommand{\bF}{{\bf F}}
\newcommand{\sF}{{\cal F}}
\newcommand{\sS}{{\cal S}}
\newcommand{\sC}{\bar{C}}
\newcommand{\dC}{\delta C}
\newcommand{\sL}{{\cal L}}
\newcommand{\sO}{{\cal O}}

\newcommand{\bO}{{\bar{O}}}

\begin{document}

\title{EXACT SHORT TIME DYNAMICS FOR STEEPLY REPULSIVE
POTENTIALS}
\author{James W. Dufty}
\affiliation{Department of Physics, University of Florida\\
Gainesville, FL 32611}
\author{Matthieu H. Ernst}
\address{Instituut voor Theoretische Fysica, Universiteit Utrecht\\
Postbus 80.195, 3580 TD Utrecht, The Netherlands}
\date{1-25-04}
\maketitle

\begin{center}{\bf Abstract} \end{center}

The autocorrelation functions for the force on a particle, the
velocity of a particle, and the transverse momentum flux are
studied for the power law potential $v(r)=\epsilon (\sigma
/r)^{\nu }$ ( soft spheres). The latter two correlation functions
characterize the Green-Kubo expressions for the self-diffusion
coefficient and shear viscosity. The short time dynamics is
calculated exactly as a function of $\nu $. The dynamics is
characterized by a
universal scaling function $S(\tau )$, where $\tau =t/\tau _{\nu }$ and $%
\tau _{\nu }$ is the mean time to traverse the core of the
potential divided by $\nu $. In the limit of asymptotically large
$\nu $ this scaling function leads to delta function in time
contributions in the correlation functions for the force and momentum
flux. It is shown that this singular limit agrees with the
special Green-Kubo representation for hard sphere transport
coefficients. The domain of the scaling law is investigated by
comparison with recent results from molecular dynamics simulation
for this potential.\\

\renewcommand{\theequation}{I.\arabic{equation}}
\setcounter{section}{0} \setcounter{equation}{0}
\section{Introduction}

The dynamics of fluctuations in simple classical fluids is a
well-studied problem over a wide range of densities and
temperatures. Important questions remain open at the quantitative
level, but important qualitative features (e.g. behavior at high
density, long times) have been largely resolved over the past few
decades. In most cases the qualitative features do not depend
sensitively on the form of the pair potential for interactions
among the particles. An exception is the short time behavior of
time correlation functions where the dynamics does depend
sensitively on the form of the potential. This is due to the
dominance of trajectories of pairs of particles as they traverse
their common force field on this time scale. The most striking
example of this is the difference between a continuous potential
with a finite pair interaction time  and hard spheres, for which
this time is zero. The objective here is to explore this
difference quantitatively for the case of time correlation
functions characterizing the Green-Kubo transport coefficients.

This work is motivated by the recent series of molecular dynamics
studies of the same problem by Powles and co-workers \cite
{Powles1,Powles2,Powles3,Powles4}. They consider a steeply
repulsive potential of the form $v(r)=\epsilon (\sigma /r)^{\nu}$
with exponent $ 12\leq \nu \leq 1152$. In the following this will
be referred to as the soft sphere potential. Clearly, for
asymptotically large $\nu $ this approaches the hard sphere
potential  which is infinite for $r<\sigma $ and zero for
$r>\sigma $ (see Figure 1).
\begin{figure}[t]
\includegraphics[width=0.6\columnwidth]{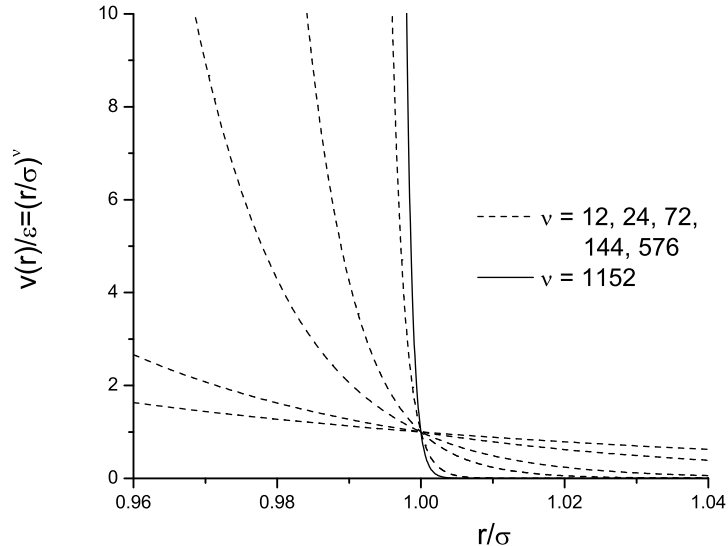} 
\caption{Soft sphere potential $v(r)$ as a function of $r/\sigma$ 
for several values of$\nu$.} 
\end{figure}

It is straightforward to show that the thermodynamic and
structural properties of the soft sphere fluid are continuously
related to those of the hard sphere fluid, as expected. However,
the corresponding relationship for dynamical properties is more
complex. For example, the exact short time expansion of time
correlation functions for the soft sphere fluid is a series with
only even coefficients \cite{deGennes}, while that for the hard
sphere fluid has finite odd order contributions as well
\cite{Longuet-Higgins}. A second qualitative difference is the
form of the Green-Kubo expressions for transport coefficients. For
the soft sphere fluid these are time integrals of the flux
autocorrelation function, with the flux being associated with some
conserved density. For the hard sphere fluid it looks as if there
is an additional term due to instantaneous momentum transport
caused by configurations for pairs of particles initially at
contact. Nevertheless, it is clear that the soft sphere and hard
sphere fluids should be physically equivalent for large $\nu $.
So, such apparent differences must be understood in an appropriate
context.

The context refers to the time scale on which the comparison is to
be made. For the soft sphere potential there is a characteristic
force range $r_{\nu }=$ $ \sigma /\nu $ \ around $r \simeq
\sigma$, and an associated  time $\tau _{\nu }=\sqrt{\beta m}
\sigma /\nu $, which is essentially the time it takes a particle
to traverse the steep part of the potential. For large $\nu $ and
$t>\tau _{\nu }$ a pair of particles initially separated by
$r\approx \sigma $ will have transferred an amount of momentum
quantitatively approaching that for a pair of hard spheres. On
this time scale all dynamical properties of the soft sphere and
hard sphere fluid should be comparable. However, for $t<\tau _{\nu
}$ the soft sphere and hard sphere fluid are always qualitatively
different in their dynamics regardless of how large $\nu $ is
taken. In effect the hard sphere fluid is always in the domain of
times large compared to $\tau_\nu$, as the collisions are
instantaneous. This explains, for example, the above difference in
the two short time power series expansions. One of the main
results described here is an exact determination of the crossover
behavior for the soft sphere fluid from $t<\tau _{\nu }$ to the
hard sphere form for $t>\tau _{\nu }$ for large $\nu .$

The detailed analysis given here is made possible by the
simplifications that occur  for repulsive power law potentials at
large $\nu $. This entails both a limitation to times $ t\approx
\tau _{\nu }$ and restricted spatial domains $\sigma -r_{\nu
}<r<\sigma +r_{\nu }$. The relevant other time and space scales
are the mean free time $t_E$ and the mean free path $l_{E}$.
Although both of these can be quite small at high densities (e.g.,
$l_{E}<\sigma $) they are insensitive to $\nu $. This means that
for sufficiently large $\nu $ there is a separation of both time
and space scales, $\tau _{\nu }<< t_{E}$ and $r_{\nu } <<l_{E}$.
As a consequence the dynamics to be studied reduces to pair
dynamics since force ranges of different pairs will not overlap
and times are much shorter than that for sequences of pair
collisions. The problem of evaluating the pair dynamics on this
time scale has been solved for the velocity autocorrelation
function (VACF) for three-dimensional soft spheres by de Schepper
\cite{deSchepper}. His analysis can be extended to other time
correlation functions and different dimensions as well. It is
found that the time domain for $t<< t_{E}$ and $\tau_{\nu } <<
t_{E}$ is described by a universal scaling function $\sS(\tau )$
depending on time only through $\tau = t/\tau _{\nu }$ and
otherwise independent of density, temperature, and potential
parameters. As it turns out, the time dependence of all Green
Kubo correlation functions are related to a single scaling
functions $\sS(\tau )$. The domain of validity for this scaling
law is studied by comparison with the three-dimensional simulation
results of Powles et al. at $\nu = 1152$ and at a packing fraction
of $\xi =0.3$. A more complete comparison at $\nu = 1152$, at
higher densities, and for more general time correlation functions
will be given elsewhere \cite{ErnstDufty}.

The initial values of correlation functions of fluxes involving
the force are proportional to $\nu $ for large $\nu $
\cite{Zwanzig,Heyes}. The combination of $\nu S(t/\tau _{\nu })$
becomes proportional to a Dirac delta function in $t$ in the limit
$\nu \rightarrow \infty $ at fixed $t$. This is the domain $
t>\tau _{\nu }$ for which the hard sphere limit is expected. To
explore the crossover to hard sphere behavior and the implications
of this delta function a detailed description of the Green-Kubo
relations and time correlation functions for the hard sphere fluid
are included here as well. The dynamics for hard spheres is no
longer described by forces. Instead there are straight line
trajectories for all particles until a pair is in contact.
Instantaneously, the pair exchanges momentum according to an
elastic collision and proceeds along the new straight line
trajectory. The generators for this dynamics (Liouville operators)
involve binary collision operators rather than forces \cite
{ErnstHS}. These differences from the dynamics for continuous
potentials lead to the qualitative differences in the Green-Kubo
relations and associated correlation functions mentioned above.
Generically, the time correlation functions $C(t)$ approach in the
limit as $\nu \to \infty$ to $ \sC (t) = \sL_\infty \delta^+
(t) + \dC (t)$. The first term represents a singular part,
which is delta-correlated in time, and a regular smooth function
$\dC (t)$. Here $\delta ^{+}(t)$ is the delta function normalized
to unity over the positive time axis. The coefficient, ${\cal L}
_{\infty }$, vanishes for any continuous potential but is non-zero
for hard spheres. Consequently, the Green-Kubo formulas for the
transport coefficients, $\sL$, in the hard sphere limit have the
form, $ \sL
 = {\cal L}_{\infty }+\lim_{t\rightarrow \infty }\lim_{V \to
\infty }\int_{0}^{t}ds \dC(s)$.

This limiting behavior is confirmed analytically below and shown
to be consistent with the simulation data. Although the literature
on hard sphere fluids is large, the detailed forms for hard sphere
Green-Kubo relations and discussion of these differences do not
seem to have been given before. Instead, an equivalent form of
Helfand relations has been used in the literature both for
theoretical analysis \cite{Helfand,DEC-helfand,Chaos-Dorf-Gasp},
and for computer simulations \cite{Allen}, since this form does
not involve the forces explicitly, and is valid for both the soft
sphere and the hard sphere fluid. Here the Helfand relations are
taken as the starting point for derivation of the hard sphere
Green-Kubo relations.

A preliminary report on the issues considered here has been given
by one of us \cite{Dufty}, restricted to the case of the shear
viscosity and without calculating the scaling function. There is a
substantial literature on the dynamics for continuous potentials
on time scales comparable to  time $\tau_\nu$, typically
associated with memory function models \cite{Boon}. This
domain is important for many conditions in neutron scattering,
spectroscopy, and short-pulse laser experiments where high
frequency domains can be accessed. The attention here is more
focused on the limiting form of the dynamics for the soft sphere
potential and the associated universal scaling properties. For a
given potential, e.g. Lennard-Jones, the short time dependence
will be potential specific and the results obtained here have
little or no relationship to more general potential forms.

The analysis here is exact in the limits considered. The objective
here is to clarify the differences mentioned above for soft sphere
and hard sphere interactions regarding the dynamics of the
associated correlation functions. To allow this broad scope of
discussion, attention is limited to the auto correlation functions
for the force, which exposes most directly the scaling function at
short times, and those for the shear viscosity and self-diffusion
coefficient. A similar analysis of the thermal conductivity and
bulk viscosity has been carried out and will be present separately
\cite{ErnstDufty}. The plan of the paper is as follows. In the
next section the Green-Kubo relations for the shear viscosity and
self-diffusion coefficient are recalled for the case of smooth
(differentiable) potentials. The fluxes defining the associated
correlation functions are identified as the sum of kinetic part
(k) and a collisional transfer or potential part (v), and the
correlation functions are decomposed into the corresponding
contributions from these components. Also in this section the
equivalent Einstein-Helfand formulas are recalled. These formulas
are applicable to both smooth interactions and hard spheres. The
generators for hard sphere dynamics are given in Appendix A and
applied in the Helfand formulas to derive the Green-Kubo formulas
for hard spheres discussed in Section III. For later comparison
with the limiting forms for the soft sphere fluid, the short time
behavior of the hard sphere correlation functions is calculated in
Section IV. The short time dynamics of the soft sphere fluid is
addressed in Section V, where the scaling properties of the
velocity autocorrelation function and force autocorrelation
function are discussed. The dominant short time contribution to
the autocorrelation function for the shear viscosity, which is the
stress autocorrelation function, is given by the same scaling
function. These results are then compared with the MD simulation
results of Powles et. al. in Section VI and final comments are
offered in the last section.

\renewcommand{\theequation}{II.\arabic{equation}}
\setcounter{section}{1} \setcounter{equation}{0}
\section{Time Correlation Functions for Smooth Interactions }

Our goal is to study the short time behavior of time
correlation functions $C(t)= (\beta/V)\avo{J(0) J(t)}$ of
microscopic fluxes $J$ that enter in the Green-Kubo formulas for
the transport coefficients in classical fluids with smooth
interactions. These fluxes contain in general a kinetic
part, $ J^k$, and a potential part,
$J^v$, i.e.  $J= J^k+ J^v$. In this paper we only illustrate the
general results by considering the most simple cases, being the
velocity autocorrelation function (VACF), the force
autocorrelation function (FACF), and the stress autocorrelation
function (SACF). The time integrals over the VACF and the SACF
determine the self-diffusion coefficient and shear
viscosity, respectively.

For smooth interactions the time correlation functions are regular
at the origin, i.e. they can be expanded in powers of $t^2$. In
the limit where the repulsive part of the potential becomes very
steep, and approaches the hard sphere limit, singularities develop
at $t=0$. The type of singularities is quite different for the
different $ab-$parts of the correlation functions,
$C^{ab}(t)=(\beta/V)\avo{J^a(0) J^b(t)}$  with
$ab=\{kk,kv,vk,vv\}$.

We start with the Green-Kubo formula for  shear viscosity
expressed in the grand canonical ensemble, characterized by a
temperature $T=1/k_B \beta$, a chemical potential, and a volume
$V$. Here averages are denoted by $\avo{\cdots}$. Then,
\be \Label{2.1}
\eta = \lim_{t\to \infty} \lim_{V \to \infty} \eta_V(t) =
\lim_{t\to \infty} \lim_{V \to \infty}\int_0^t ds C_\eta(s),
\ee
where $ \eta_{V}(t)$ is a time integral over the time
correlation function,
\be\Label{2.2}
C_\eta(t) = (\beta/V)\avo{J(0) J(t)}=C_\eta^{kk}(t)+2
C_\eta^{kv}(t)+C_\eta^{vv}(t).
\ee
Here  $C_\eta^{kv}(t)=C_\eta^{vk}(t)$ are equal, and the fluxes
are given by
\ba \Label{2.3}
J &=& \sum_i mv_{ix} v_{iy} +\sum_{i<j} r_{ij,x}F_{ij,y}
\nn J^k &=& \sum_i mv_{ix}v_{iy}; \qquad
 J^v =\sum_{i<j} r_{ij,x}F_{ij,y}.
\ea
Similarly, the Green-Kubo formula for the self diffusion
coefficient $D$ is given by
\be \Label{2.4}
D=\lim_{t\to \infty} \lim_{V \to \infty}D_V(t) = \lim_{t\to
\infty} \lim_{V \to \infty}\int^t_0 ds C_D(s),
\ee
where $C_D(t)$ is the autocorrrelation function for the velocity
of one of the particles,
\be \Label{2.7}
C_D(t)= \avo{v_{1x}(0) v_{1x}(t)}.
\ee
A self-correlation in the grand canonical ensemble has to be
understood as $\avo{a_1(0)a_1(t)} = \avo{\sum_i
a_i(0)a_i(t)}/\avo{N}$. Finally, we will also consider here the
autocorrelation function for the total force acting on a single
particle
\be \Label{2.7a}
C_F(t)= (\beta/m)\avo{F_{1x}(0) F_{1x}(t)}.
\ee
where $F_{1x}=\sum_{i>1} F_{1i,x}$ is the force on particle $1$.
The FACF is simply related to the second derivative of the VACF.
Its time integral does not give any transport coefficient, but
vanishes in fact. However, it is included in this study since it
has a time dependence closely related to that of the SACF. For
simplicity it will be referred to as a "Green-Kubo correlation
function" along with $C_D(t)$ and $C_\eta(t)$.

Next we consider the Einstein-Helfand Formulas. These formulas
\cite{Helfand,Allen} for the transport coefficients are the
analogs of the Einstein formula for the self diffusion coefficient
$D$ in terms of the second moment of the displacements,
\be \Label{2.8}
D_V(t) =\half \frac{d}{dt} \avo{(x(t)-x(0))^2}= \int^t_0 ds
\avo{v_{1x}(0)v_{1x}(s)}.
\ee
The first equality is the Einstein-Helfand form while the second
equality is the Green-Kubo form. The latter follow directly from
the identities $\dot{x} =v_x$ and $x(t)-x(0) = \int^t_0 ds
\dot{x}(s)$, and the stationarity of the VACF. The Helfand formula
for the shear viscosity is given by an analogous moment formula,
\ba \Label{2.9}
\eta_V(t) &=&\frac{\beta}{2V} \frac{d}{dt} \avo{(M(t)-M(0))^2}
\nn M &=& \sum_i m v_{ix}r_{iy},
\ea
where $M$ is a Helfand moment. The equivalence of \Ref{2.9} with
the Green-Kubo formula can be established along the same lines as
in \Ref{2.8} by observing that the flux is given by,
\be \Label{2.11}
J =\dot{M} =\{M,H\} \equiv L M,
\ee
where $H=K+V$ is the Hamiltonian, and $\{M,H\}$ is a Poisson
bracket. The last equality defines the Liouville operator for
smooth potentials. Expressions in terms of a Liouville operator
are of interest as we want to consider also interactions between
hard spheres. Such interactions can not be described by a
Hamiltonian being a smooth function of the relative distances
between the particles. However, there exist in the literature
pseudo-Liouville operators, that generate the hard sphere dynamics
inside statistical averages \cite{ErnstHS}. The Helfand formulas
are given in terms of the moments $M$ which do not involve the
force that becomes singular for the hard sphere limit, in
contrast to the fluxes in the Green-Kubo representation. As
indicated in the next section, use of the hard sphere
pseudo-Liouville operators in \Ref{2.11} provides the correct
definition for the microscopic hard sphere fluxes and the
corresponding hard sphere Green-Kubo representation using
\Ref{2.11}.

\renewcommand{\theequation}{III.\arabic{equation}}
\setcounter{section}{2} \setcounter{equation}{0}
\section{Time Correlation functions for hard spheres fluids }
\subsection{Einstein-Helfand and Green-Kubo Formulas}

The corresponding time correlation functions for hard sphere
fluids, denoted by $\sC_\mu^{ab}(t)$, are singular at $t=0$. They
behave quite differently at short times from those for smooth
interactions, which are regular at $t=0$. For instance, the VACF
for hard spheres decays exponentially on the time scale of the
Enskog mean free time $t_E$, and it is {\it even} in $t$. So, it
is singular at $t=0$ with a jump in the first derivative, and a
delta function in the second derivative.

In order to investigate what happens in the limit of very steep
repulsion, we want to first calculate the corresponding  hard
sphere results, which have only been studied in the literature for
the VACF and the incoherent scattering function
\cite{deSchepper,dSEC}. The problem is that the fluxes $ J$ in
\Ref{2.3} contain the forces $\bF_{ij}$, which are ill-defined for
hard spheres, and there is no obvious way to extend the usual
Green-Kubo formulas to hard sphere fluids. However, as noted
above, the equivalent Einstein-Helfand formula \Ref{2.9} involve
only momenta and energies which remain well-defined in the hard
sphere limit.

The proper way to formulate hard sphere dynamics for an
equilibrium time correlation function $\avo{ A(0) B(t)}$ is
to introduce the forward (+) and backward (-) generators $e^{\pm t
L_{\pm}}$ with $t>0$. They generate the trajectories in $\Gamma-$
space, $A(t) = e^{\pm t L_{\pm}}A(0)$, for an arbitrary phase
function inside a statistical average. The generators involve the
pseudo-Liouville operators $L_\pm$ and the binary collision
operators $T_\pm$ defined in Appendix A (see Ref \cite{ErnstHS}),
where we also explain how to express the Einstein-Helfand
correlation function for $\eta$ in terms of these hard sphere
generators, and subsequently into a Green-Kubo formula. The result
is,
\be \Label{3.1}
\eta_V(t) = \int_0^t ds \,\sC_\eta (s)= \eta_\infty + \int_0^t ds
\, \dC_\eta (s),
\ee
where the limits in \Ref{2.1} still have to be taken. Here we have
introduced,
\ba \Label{3.2}
{\sC}_\eta(t) &=& \eta_{\infty} \delta^+(t) + \dC_\eta(t); \qquad
{\dC}_\eta (t) = (\beta/V) \avo{J_{-} e^{tL_+}  J_{+}}
\nn  J_{\pm} &=& J^k + J^v_{\pm} = L_\pm M ;\qquad
J^k = L_0 M= \sum_i mv_{ix} v_{iy}
\nn  J^v_{\pm} &=&  \pm \sum_{i<j} T_\pm(ij)M =
 \pm \sum_{i<j} T_\pm(ij) \half m g_{ij,x}r_{ij,y}
\nn \eta_{\infty} &=& -(\beta/V) \avo{M L_+ M}.
\ea
The delta-function is normalized as $\int_0^\infty dt
\delta^+(t)=1$. The kinetic part of the flux in \Ref{3.2} is
identical to that for smooth interactions \Ref{2.3}. For the time
being the collisional part, involving $T-$operators, will be kept
in the schematic form above.

Next we consider the instantaneous viscosity $\eta_\infty$, which
is defined as an equilibrium average, and vanishes for smooth
interactions ( i.e., when $L_+$ is replaced by $L$). For the hard
sphere fluid $\eta_\infty$ can be expressed in terms of the hard
sphere pair distribution function by combining the expressions for
$\eta_\infty$ and $L_+M= J_+$ in \Ref{3.2}, where we have changed
to center of mass, $\{ {\bf R_{ij},G_{ij} =\half(\bv_i +
\bv_j})\}$, and relative phase variables, $\{{\bf r_{ij},
g_{ij}=\bv_i-\bv_j}\}$. The result is
\be \Label{3.6a}
\eta_\infty =-\eighth \beta(mn)^2 \chi \int d\br \dav{g_xr_y
 \; T_+(12)\;g_x r_y} ,
\ee
where $ \dav{\cdots}$ denotes a Maxwellian velocity average over
all particles involved. Moreover, the $\br-$integration can be
carried out because the operator $T_+$ contains a factor
$\delta^{(d)}(\br - \sigma {\bf \hsig})$. Consequently $\chi
=g^{(2)}(\sigma +)$ is the hard sphere pair correlation function
at contact. The remaining integrals are $d-$dimensional
generalizations of the collision integrals as appearing in the
Enskog theory for hard sphere fluids (See Chapter 16.8 of Ref.
\cite{Chapman}). Performing the $ \hbsig-$ and velocity
integrations yields finally,
\be \Label{3.6}
\eta_\infty = \frac{m n \sigma^2}{d(d+2) t_E} \equiv \frac{d}{d+2}
\varpi.
\ee
The Doric pi $\varpi$, defined in \Ref{3.6}, has been chosen such
that it reduces for $d=3$ to the same symbol as used in the
classical Enskog theory, as presented in Chapters 16.51, 16.52 and
16.6. of Ref. \cite{Chapman}. In the above formula the Enskog mean
free time $t_E$ is given by,
\be \Label{3.7}
t_E =\sqrt{\pi} t_\sigma / 2d b n \chi \equiv \sqrt{\pi} t_\sigma
/ 2d \Delta
\ee
where $t_\sigma =\sqrt{\beta m}\sigma $ and $ b$ is the excluded
volume, equal to half the volume of the $d$-dimensional
interaction sphere of radius $\sigma$.

There are two conspicuous differences  between the formulas for
hard spheres and for smooth interactions. First, the
pseudo-fluxes $J_{\pm}$ in the time correlation functions are
different depending on their position relative to the generator
$e^{tL_+}$. We also note that $ J_+e^{-tL_-} J_-$ is an equivalent
order. Second, there is the instantaneous contributions,
$\eta_{\infty} $, which is vanishing for {\it smooth }
interactions.

Before closing this section we point out that the hard sphere time
correlation formula $\sC_\eta(t)$ in \Ref{3.2} can also be split
into $\sC_\eta^{ab}(t)$ with $(ab) =\{kk,kv,vv\}$ by splitting the
flux as $ J_{\pm} =  J^k+ J^v_{\pm}$ (see \Ref{3.2}). The form of
the kinetic part $J^k$ is identical to that for smooth
interactions. The remaining collisional transfer part $ J^v_{\pm}$
is different.

\subsection{ The hard sphere FACF and VACF}

In the same way as described above for $C_\eta (s)$, the Helfand
representation for the force autocorrelation function can be used
to obtain the equivalent hard sphere form
\ba \Label{AB}
{\sC}_F(t) &=& \gamma_\infty \delta^+(t) + \dC_F(t)
\nn \dC_F(t) &=& \frac{\beta}{m} \avo{F_{1x-} e^{tL_+}F_{1x+}}
\nn  F_{1x,\pm} &=& L_\pm mv_{1x} = \pm \sum_{j \neq1}T_\pm(1j)  mv_{1x}.
\nn \gamma_\infty &=& - \beta m \avo{v_{1x}L_+v_{1x}}= 4 \Delta /
 \sqrt{\pi} t_\sigma =2/d t_E.
\ea
One directly recognizes the expression for $\gamma_\infty$ as the
opposite of the initial slope of the hard sphere VACF, which has
been calculated exactly along the same lines as \Ref{3.6} (see
Ref.\cite{dSEC}). We also note that $\gamma_\infty$ vanishes for
smooth interactions.

Finally, the VACF for hard spheres is much simpler to obtain since
there is no explicit dependence on the forces. Consequently, only
the generator for the dynamics has to be changed, leading to
\be \Label{3.10}
{\sC}_D(t)= \avo{v_{1x}(0) e^{tL_+}v_{1x}}
\ee

This completes our identification of the hard sphere Green-Kubo
time correlation functions $\sC_\eta (t)$, $\sC_F (t)$,
and $\sC_D (t)$.

\subsection{ A new relation for the hard sphere fluid}

An interesting consequence for the hard sphere FACF follows by
explicitly integrating \Ref{2.7a}  over $t$ using Newton's law to
find for, say, soft spheres,
\be \Label{3.12}
\lim_{t \to \infty} \int^t_0 ds C_F(s) = \beta \lim_{t \to \infty}
\avo{F_x(0)(v_{1x}(t)- v_{1x}(0))} =0,
\ee
because positions and velocities become uncorrelated in this
limit. Taking the limit of the exponent of the power law
interaction $\nu \to \infty$ suggests that this relation  also
holds in the case of the hard sphere correlation function $\sC
(t)$ in \Ref{AB}, i.e. $\int_0^\infty ds \sC_\eta(s)=0$. As a
consequence we obtain the following relation,
\be \Label{3.14}
\gamma_\infty  = - \frac{\beta}{m} \int^\infty_0 dt
\avo{F_{1x-}e^{tL_+}F_{1x+}}.
\ee
The left side of the last equation has been calculated exactly
above. The right side is a complex dynamical quantity involving
the entire time evolution of the system. Thus we have obtained a
rare "zero frequency" sum rule for the hard sphere fluid. It
should be noted however that equating \Ref{3.12} with the
corresponding integral over $\sC$, implies an interchange of
limits, i.e. $\lim_{t \to \infty}$ and $\lim_{\nu \to \infty}$,
which is presumably allowed. However, interchanging the limits,
$\lim_{t \to 0}$ and $\lim_{\nu \to \infty}$, is not allowed, and
the consequences of this non-uniformity is in fact the main
subject of this paper. The full implications of the new relation
\Ref{3.14} are not clear at this point.

\renewcommand{\theequation}{IV.\arabic{equation}}
\setcounter{section}{3} \setcounter{equation}{0}
\section{Evaluation of Hard Sphere Properties}

\rm Having obtained Green-Kubo correlation functions
$C_\mu^{ab}(t)$ for smooth interactions  and $\sC_\mu^{ab}(t)$ for
hard sphere interactions, we now discuss the structure of the
short time behavior of these functions. After a brief introduction
we consider in this section the hard sphere case, and in Section V
the case of smooth interactions, in particular soft spheres.

First, some perspective on the differences between these two cases
is given by listing in Table I the qualitative structure of the
short time behavior for smooth and hard sphere interactions. We
first note that all Green-Kubo-type correlation functions are even
functions of time. If they are regular at $t=0$, as is the case
for smooth interactions, then they can be expanded in powers of
$t^2$. For hard sphere systems the correlation functions are also
even functions of $t$, but they are singular at the origin.

\begin{center} \sc Short time behavior of time correlation functions\\
for smooth interactions and for hard sphere interactions.
\[\begin{array}{|c|c|c|c|c|c|c|} \hline \hline
\mu \downarrow &\multicolumn{3}{c|}{C^{ab}_\mu(t) \mbox{ for SI}}&
\multicolumn{3}{c|}{\sC^{ab}_\mu(t) \mbox{ for hard sphere}}
\\ \hline ab \to & kk & kv &vv &kk &kv & vv
\\ \hline D & O(1)+O(t^2) &  && O(1)+\bO(|t|) &  &
\\  F    & & & O(1)& & & \delta (t)
\\  \eta & O(1)+O(t^2) & O(t^2)& \;O(1)\;& O(1)+\bO(|t|) & \;\bO(1)\;& \;\delta (t)\;
\\ \hline \hline
\end{array}\]
\end{center}

Table I shows the schematic structure of the selected  correlation
functions for smooth interactions and hard sphere fluids, where
the terms $O(t^n)$ and $\bO(t^n)$ with $n=0,1,2$ are non-vanishing
terms of order $t^n$ for small $t$. The results for smooth
interactions are well-known in the literature. Regarding hard
spheres we first note that at small $t$ only the leading order
terms (initial values) in $C^{kk}_\mu(t)$ and
$\sC^{kk}_\mu(t)$ are equal. The entries in the remaining columns
for smooth interactions and hard spheres are all different. The
hard sphere results are obtained from \Ref{3.2} for $\sC_\eta
(t)$, \Ref{AB} for $\sC_F (t)$, and \Ref{3.10} for $\sC_D (t)$.
All contributions involving a single $T-$operator are
non-vanishing, as in the hard sphere entries on location
$(\mu,kk)$ and $(\mu,kv)$. The hard sphere entries on location
$(\mu,vv)$ involve two $T-$operators, and are more complicated.
They will be discussed later.

Inspection of Table I shows that the results for smooth
interactions and hard spheres  are indeed very different. The goal
of this section and the next is to calculate the entries in the
table, and study in a quantitative manner the crossover of the
correlation functions to hard sphere interactions from smooth, but
steeply repulsive power law interactions, $ v(r) \sim 1/r^\nu$.
This will be done for times $t$, short compared to the Enskog mean
free time $t_E$.

In the remaining part of this section the short time behavior of
the correlation functions for the hard sphere fluid in Table I
will be calculated, i.e. the initial values and initial slopes. We
start with the (kk)-correlations, and include the VACF $\sC_D(t)$
as the most typical one. In the sequel we restrict ourselves
exclusively to $t>0$, to avoid possible confusion regarding the
definitions of the $T-$operators.
\ba \Label{6.4}
\sC_D (t) &=& \avo{v_{1x}\{1+tL_+ \cdots \}v_{1x}} \equiv (1/\beta
m)\{ 1-\gamma_D t+\cdots \}
\nn \sC^{kk}_\eta (t) &=& \avo{J^k\{1+tL_+ \cdots \}J^k} \equiv (n/\beta)\{
1-\gamma_{\eta} t+\cdots \}.
\ea
As explained in the previous subsection, we restrict ourselves in
the small-$t$ expansion to terms that are at most linear in the
$T-$operator. Terms of $\sO ((tT)^2)$ have been neglected in
\Ref{6.4}. Initial values and slopes can be evaluated, and yield
\ba \Label{6.6a}
\gamma_D &=& \gamma_\infty= -\avo{v_{x}L_+ v_{x}}/ \avo{v_x^2}=
2/d t_E \nn \gamma_\eta &=& -\avo{J^k L_+ J^k}/\avo{(J^k)^2}=
{4}/(d+2)t_E
\ea
with $J^k$ defined in \Ref{2.3}.

Next we consider the (kv)-cross-correlations $\sC^{kv}_\eta (t) =
(\beta/V) \avo{ J^k \exp[t L_+]J^v_+}$. As $J^v_+ \equiv J_+ -
J^k$ in \Ref{3.2} itself is already linear in $T$, we can only
calculate its initial value exactly,
\ba \Label{6.14}
& \sC^{kv}_\eta (t) = (\beta/V) \avo{J^k \{ 1+ \cdots\} \sum_{i<j}
T_+(ij) \half m g_{ij,x} r_{ij,y}}&
\nn &= \eighth \beta(mn)^2 \chi  \int d \br \dav{
g_x g_y T_+(12)g_x r_y} =\left(\frac{n}{\beta}\right) \frac{2
\Delta}{d+2}&.
\ea
Finally we discuss the (vv)- or collisional transfer correlations
for hard spheres in Table I. It is instructive to first
compare $C^{vv}_\eta (t)$ for smooth power law interactions with
$\sC^{vv}_\eta (t)$ for hard spheres. The correlation function
$C_\eta^{vv} (t)$ for smooth interactions develops in the hard
sphere limit a strong delta function - type singularity, as
represented by the first line in \Ref{3.2} -- and similarly in
\Ref{AB} for $C_F(t)$. The remaining part, $\dC_\eta^{vv} (t)$,
containing the pseudo-fluxes $J^v_\pm$, represents in fact the
regular part, that approaches a finite limit as $t \to 0$. Indeed,
the short time behavior of $\dC_F(t)$, or equivalently
$\ddot{\sC}_D(t)$, has already been analyzed in great detail in
the literature (see Ref.\cite{dSEC}). There it has been shown that
the hard sphere correlation functions related to
$\delta\ddot{C}_D(t)$ or $\dC_F(t)$, and containing two
$T-$operators, are smooth functions of time near $t=0$, which
indeed approach a finite non-vanishing value. The explicit
evaluation of these contributions is much more complex than
performing simple binary collision integrals, as we have been
doing in the previous sections. The reason is that these
contributions are coming from the overlapping part of uncorrelated
binary collisions (12)(13), and from renormalized ring collisions
of the form (12)(13)(23) \cite{dSEC}, which involve in fact {\it
three} instead of two hard spheres. However, for the purpose of
this paper the fact that these values are finite, is sufficient.

\renewcommand{\theequation}{V.\arabic{equation}}
\setcounter{section}{4} \setcounter{equation}{0}
\section{Scaling forms in soft sphere fluids}
\subsection{Force autocorrelation function (FACF)}

In the previous section we have calculated the leading short time
hard sphere results. These are needed to compare and identify the
limiting results for soft sphere correlation functions on time
scales that describe the crossover to hard sphere behavior, i.e.
the crossover from the {\it initial} time scale, where the
detailed shape of the interparticle interaction matters,  to the
{\it kinetic} time scale $t_E$, where
only asymptotic scattering properties matter.\\

This will be done for {\it soft spheres}, represented by the
repulsive power law potential $v(r)= \epsilon (\sig/r)^\nu$. In
this case the FACF and SACF turn out to be proportional to a
scaling function ${\cal S}( t/\tau_\nu)$, where $\tau_\nu =
t_\sig/\nu$ is the mean time that a particle needs to traverse the
steep part of the potential, and $t_\sigma= \sqrt{\beta m} \sigma$
the mean time to traverse the total hard core diameter $\sigma$.
These time scales are only well-defined and well-separated for
large values of exponent $\nu$. At high densities there is another
relevant time scale, the mean free time between collisions, $t_E$,
which can be estimated for sufficiently steep repulsion by the
Enskog mean free time, $t_E$, which is proportional to $t_\sigma /
bn \chi$. Given the large values of the two- and three-dimensional
pair correlation function $\chi$ at contact, $t_E$ and $t_\sigma$
can be of the same magnitude (for example in two- and
three-dimensional hard sphere systems at packing fractions around
30\%), or $t_E$ may even be an order of magnitude smaller than
$t_\sigma$ (for example in typical neutron scattering experiments
on liquid Argon). Hence, to describe the  crossover from the
initial $\tau_\nu-$scale to the kinetic scale $t_E$, the initial
scale must satisfy $\tau_\nu \ll \min\{t_E,t_\sigma\}$. These
estimates suggest that the exponent $\nu$ should be rather large
at high densities. The example of the next section with packing
fraction $0.3$ and $\nu=1152$ satisfies these constraints, as
will be illustrated in Section VII.

Our analysis begins with the FACF in a $d$-dimensional system,
defined in \Ref{2.7a}. It is an even function of time $t$ and
regular at the origin for finite values of the exponent $\nu$ ,
i.e. it can be expanded in powers of $t^2$. However for very large
$\nu$, the initial value  $C_F(0) \to \infty$, implying that the
function is singular at the origin. It is the goal of this
subsection to analyze the dominant small-$t$ singularity of
$C_F(t)$ in the hard sphere limit and to describe the crossover of
the FACF from soft to hard sphere behavior.

A study of this short time crossover problem can be carried out
following the work of de Schepper \cite{deSchepper} for the VACF.
His analysis is based on a perturbation expansion of $e^{tL}$ in
powers of $(tL)^k$, where the Liouville operator contains the
force $\bF_{ij}$. He has shown that the most dominant
contributions for large $\nu$ are obtained by keeping in each
order in $(tL)^k$ only terms involving forces between a single
pair $(ij)$, and finally resumming these contributions. Here we
exploit this result and calculate directly the entire pair
contribution. The basic physical idea is that the autocorrelation
function $\avo{\bF_{12}(0)\bF_{12}(t)}$ of the pair force
$\bF_{12} = - {\bf \nabla}v(r) \sim \sO(\nu)$ controls the short
time dynamics on the time scale $\tau_\nu$, and the time evolution
is controlled by two-particle dynamics since $ \tau_\nu \ll t_E$.
More explicitly, the dominant short time contribution to $C_F(t)$
for $\nu \to \infty$ is,
\ba \Label{4.2}
C_F(t) &\simeq & (\beta n/ m V) \int d\bR d\br d\bg d{\bf G}\,
\phi (g) \phi (G) g^{(2)}(r) F_x(r) e^{tL_{12}} F_x(r)
\nn &\equiv & (\beta n \chi/ m) \sav{ F_{12,x} e^{tL_r} F_{12,x}}.
\ea
The second line defines a two-particle average over positions and
velocities, the latter one with  Maxwellian weights. A
change of variables from $\{\br_1,\bv_1,\br_2,\bv_2 \}$ to
relative and center of mass coordinates $\{\br,\bg,{\bf R,G}\}$
has been made with the replacement $L_{12}=  L_r +\bG \cdot {\bf
\nabla_\bR}$, and $n=\avo{N}/V$. For large $\nu$ the
$\br-$integrand is sharply peaked around $r=\sigma$. Consequently
the pair distribution function $g^{(2)} (r)$ can be replaced by
its value at contact, $\chi \equiv g^{(2)} (\sigma +)$.

The analysis is now reduced to a one body problem in the soft
sphere potential. The detailed calculations are still rather
technical, and will be published elsewhere. The final result is
\be \Label{4.7}
C_F(t) \simeq (\gamma_\infty /\tau_\nu) \sS (t/\tau_\nu).
\ee
where the crossover function $\sS(\tau)$ is found as,
\be \Label{4.9}
\sS(\tau) \equiv  \left( \frac{d}{d \tau}\right)^2 2 \tau
\int^\infty_0 dy e^{-y^2} y^3 \coth \tau y.
\ee
Interestingly, the crossover function is independent of the
dimensionality $d$, and  depends on the exponent $\nu$ only
through the scaling variable $\tau = \nu t/t_\sigma$. All explicit
$\nu-$dependence in \Ref{4.7} is accounted for in the overall
factor $\nu$contained in $1/\tau_\nu$.

The crossover function $\sS(\tau)$ can be expanded, both at short
and at long times, in a convergent infinite series with known
coefficients, of which we quote the leading terms,
\be \Label{4.12}
\sS(\tau) \simeq \left\{ \begin{array}{ll} \half \sqrt{\pi}
(1-\tau^2)
+{\sO}(\tau^4) &  \qquad(\tau < \pi)  \\
 \frac{\pi^4}{5}\tau^{-5} +{\cal O}(\tau^{-7}) &\qquad
(\tau > \pi)  \end{array} \right .
\ee
A numerical evaluation is shown in Figure 2.
\begin{figure}[t]
\includegraphics[width=0.48\columnwidth]{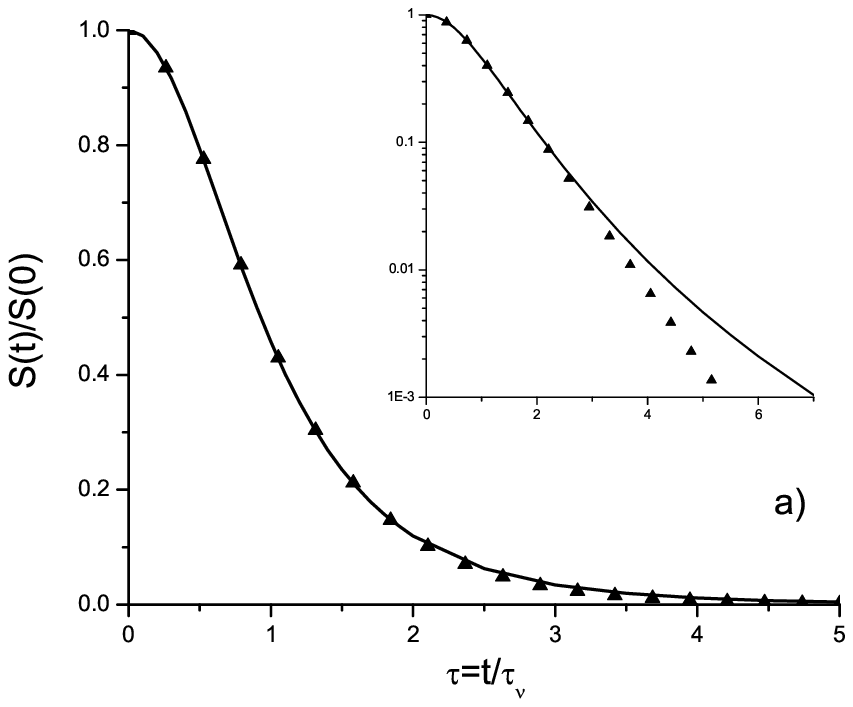}
\includegraphics[width=0.48\columnwidth]{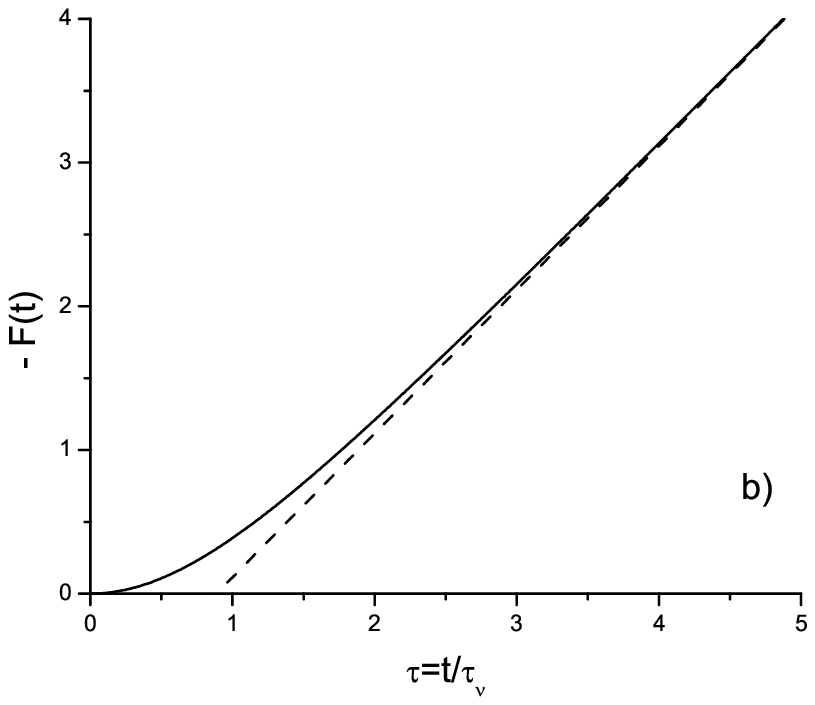}
\caption{Figure (a) shows the exact scaling function $\sS(\tau)$ (solid line) 
together with the phenomenological one $1/\cosh(\tau \sqrt{2})$ (triangles), 
as discussed in the text. The insert shows the same on a logarithmic scale, 
in order to visualize the differences between the exact and phenomenological one  
at large $\tau$.Figure (b) shows the scaling function $\sF(\tau)$ (solid line) for
the VACF together with its large$-\tau$ asymptote (dashed line).}
\end{figure}

The scaling property was first noticed in the MD simulation
results of Powles and collaborators
\cite{Powles1,Powles2,Powles3,Powles4}. In that work a
phenomenological crossover function $\propto 1/\cosh(\tau
\sqrt{2})$ has been used, which decays exponentially. The insert
in Figure 2 also shows that it is correct at short times, but
decays too fast at longer times.

The short time behavior for the soft sphere fluid (fixed, but
large $\nu$) occurs for $ \tau = t/ \tau_\nu <\!< \pi$ and
therefore is obtained from the small$-\tau$ behavior in
\Ref{4.12},
\be \Label{4.12b}
C_F(t)  = \frac{\sqrt{\pi}}{d t_E} \left( \frac{1}{\tau_\nu}
\right) \left[ 1 - ({t}/{\tau_\nu})^2  + {\cal O} (
({t}/{\tau_\nu})^4)\right].
\ee

\noindent To describe the crossover to the hard sphere fluid we
observe that
\be \Label{4.12c}
  \lim_{\nu \to \infty} \frac{1}{\tau_\nu}\sS
  \left(\frac{ t}{\tau_\nu}\right)= \delta^+(t)
\ee
can be considered as a delta function on the interval
$(0,\infty)$, i.e.
\be \Label{4.12d}
\int_0^\infty dt \,\delta^+(t) = \lim_{\nu \to
\infty}\int_0^\infty d \tau \, \sS(\tau) =1.
\ee
Its value at $t = 0$ is infinite, and vanishes  at $t \neq 0$,
because $\sS(0) \neq 0$ and $\sS(\infty)=0$. The hard sphere
behavior ($\nu \to \infty$) at {\it small, but fixed $t$} , is
obtained from the large$-\tau$ behavior of $\sS(\tau)$ in
\Ref{4.12} and \Ref{4.12c}, and yields for $t/\tau_\nu > \pi$ or
$\nu > \pi t_\sigma / t$,
\be \Label{4.12e}
C_F(t) =\frac{ 2}{d t_E} \left( \frac{1}{\tau_\nu}\right) \left(
\delta^+(\frac{t}{\tau_\nu}) + \fifth \pi^4 (\tau_\nu / t)^5 +
{\sO}((t/\tau_\nu)^{7}) \right).
\ee
The first term inside the curly brackets represents the dominant
contribution to the hard sphere FACF in \Ref{AB} at short times,
and the second term represents  the dominant correction to the
limiting result.
    The identification of the crossover function and demonstration
that it generates a delta function singularity is one of the main
results of this work. It provides the basis for understanding the
connection between the soft sphere VACF and the singular form of
the hard sphere VACF shown in \Ref{AB}. It is noted that the
coefficient of the delta function arising from \Ref{4.7} is
exactly the same as that of \Ref{AB} coming from transformation of
the Helfand formula.

\subsection{Stress Autocorrelation Function (SACF)}

Application of this analysis to the collisional transfer or
vv-parts of the correlation functions for the shear viscosity,
bulk viscosity, and thermal conductivity leads to exact results
for the short time crossover function of the form \Ref{4.7} with a
different pre-factor, but with the same scaling form $\sS(\tau)$.
In the hard sphere limit this scaling form approaches according to
\Ref{4.12c} again to a Dirac delta function. The kk- and kv-parts
of the time correlation functions at short times are less singular
than the vv-parts. So the dominant short time singularity of the
full time correlation functions is contained in the collisional
transfer (vv) terms. We only quote explicitly the result for the
stress-stress correlation function,
\be \Label{4.19}
 C_\eta (t) \simeq C_\eta^{vv} (t) \simeq (\eta_\infty /\tau_\nu)
 \sS(  t/\tau_\nu).
\ee
Again, the delta function singularity and its prefactor
associated with this result
agree exactly with the singular part of \Ref{3.2}.

\subsection{Velocity Autocorrelation Function (VACF)}

Finally we consider the VACF $C_D(t)$ for the soft sphere fluid,
as defined in \Ref{2.7}. It is regular at $t=0$ with $C_D(t)
=(1/\beta m)\{ 1+\sO(t^2)\}$ according to Table I. On the other
hand, for the hard sphere fluid $\sC_D(t) =(1/\beta m)\{1 -
\gamma_\infty |t|/t_\sigma +\cdots \}$ is linear in t at small
positive $t$. Again, computing the dominant contribution of the
pair force autocorrelation function with pair dynamics we recover
the result of \cite{deSchepper},

\be \Label{4.13}
C_D(t)= \frac{1}{\beta m} \left\{ 1+ \gamma_\infty \tau_\nu
\sF(\tau) \right\} = C_D(0) \left\{1 +\frac{4
\Delta}{\sqrt{\pi}}\sF(\tau) \right\}
\ee
with $\Delta = bn \chi$   and  $\tau = t/\tau_\nu $, and the new
crossover function is
\be \Label{4.15}
\sF(\tau) = 2 \int^\infty_0 dy e^{-y^2} y^2 (1-\tau y \coth \tau y
).
\ee
The function $\sF(\tau)$ is simple related to $\sS(\tau)$ by
\be
(d^2/d\tau^2) \sF(\tau) = - \sS (\tau).
\ee
This is consistent with the exact relationship of the VACF and the
FACF, $C_F(t) = -\beta m \ddot{C}_D (t)$. The initial conditions
on $\sF(\tau)$ as $\tau \to 0$ can be read
off from location $(D,kk)$ in Table I to be $\sF(0)=0$ and
$\sF^\prime (0)=0$, which determines the two integration
constants.

The small and large $\tau-$expansions of $\sF(\tau)$ are for
positive $\tau$,
\be \Label{4.16}
\sF(\tau) \simeq \left\{ \begin{array}{ll}
-\textstyle{\frac{1}{4}} \sqrt{\pi}\,\tau^2
+{\cal O}(\tau^4) &  \qquad(\tau < \pi)  \\
-\tau +\half \sqrt{\pi} -\frac{\pi^4}{60} \tau^{-3} +{\cal
O}(\tau^{-5}) &\qquad (\tau
> \pi)  \end{array} \right.
\ee
The short
time behavior in the soft sphere system at fixed, but large
$\nu$ is now found to be,
\be \Label{4.17}
C_D(t) \simeq \frac{1}{\beta m} \left\{1 -\Delta (t/\tau)^2
 + {\cal O}((t/\tau_\nu)^4)  \right\}.
\ee
This is the explicit form of the short time behavior of $C_D(t)$,
listed in schematic form in Table I on location $(D,kk)$ for the
smooth interactions case. Similarly we find for large $t/\tau_\nu$
behavior (with $t$ fixed, but small),
\be \Label{4.18}
C_D(t) \simeq  \frac{1}{\beta m} \left \{ 1 - (4 \Delta/
\sqrt{\pi})
 [t/\tau_\nu - \half \sqrt{\pi} \:]  +
 {\sO}(\tau_\nu^3) \right \},
\ee
where the first two terms inside the curly brackets represent the
behavior of the VACF  for hard spheres at small $t$, as given
schematically in Table I, and explicitly in \Ref{6.4}.  Note that
the dominant correction of ${\cal O}(\nu^{-1})$ to the hard sphere
result is independent of $t$, and decreases very slowly with
$\nu$.

\renewcommand{\theequation}{VII.\arabic{equation}}
\setcounter{section}{5} \setcounter{equation}{0}
\section{Comparison with Simulations }

During the past five years Powles and collaborators have carried
out a systematic study of the short time dynamics for the soft
sphere fluid by molecular dynamics simulation. They report results
at two packing fractions $0.3$ and $0.45$, corresponding to a
moderately dense fluid, for the range of exponents $\nu = 12,24,
36,72, 144, 526,$ and $1152$. This is done for the correlation
functions occurring in the Green-Kubo expressions for shear and
bulk viscosity, thermal conductivity, self-diffusion coefficient,
and the force autocorrelation function. These results provide
extensive data to test and interpret the theoretical predictions
made here. As only partial results have been given here, for the
FACF, SACF and VACF, the comparison will be limited to these
cases. Furthermore, attention will be restricted to the packing
fraction $0.3$ and $\nu =1152$. These are the best conditions for
the separation of time scales $\tau_\nu \ll \min\{t_E,t_\sigma\}$.

We first observe that our theoretical results are {\it
asymptotically exact}, i.e. $\nu$ has to be large enough such that
on the one hand the gap between $\tau_\nu$ and the minimum of
$\{t_E,t_\sigma\}$ is sufficiently large to test the predictions
for asymptotically large $\tau =t/\tau_\nu$, and on the other hand
$t$ has to be short enough, i.e. $t << t_E$, such that the
dynamics of the isolated pairs describes the full time evolution
of the $N-$particle system. An estimate at $\xi=0.3$ and $\nu =
1152$ gives: $t_E \simeq 125 \tau_\nu \simeq 0.099 t_\sigma$,
illustrating that the time scales are well separated.

It is useful to anticipate the results of the comparison between
the short time scaling behavior determined here and that from the
simulations. Since the results here are asymptotically exact,
there must be agreement in all cases where $\tau_\nu <<
\{t_E,t_\sigma\}$  and $t<<t_E$. There will then be the crossover
domain where the short time form begins to fail. However, the long
time form of the scaling functions represents the "short time"
dynamics of a hard sphere fluid (fixed small $t$, large $\nu$), as
indicated in the previous section. This is the domain for which
the VACF exhibits behavior, linear in $t$.

What happens for larger $t$, where the decay of the VACF becomes
exponential, rather than linear (say for $t > 0.5 t_E$), is
strictly outside the predictions of the present theory, and no
exact results are known for these larger time intervals. Of course
the results may still be compared with the Enskog theory for hard
sphere fluids, but we do not consider approximate results in the
present article. Similarly, the singular part of the FACF and SACF
will be very small on this kinetic time scale, and no predictions
on the regular parts of these functions, $\dC_F(\tau)$ and
$\dC_\eta (t)$, can be given in the context of the present theory.
We note that these regular parts contain, among other, the full
kk- and kv-parts of $C_\eta(t)$.
\begin{figure} 
\includegraphics[width=0.6\columnwidth]{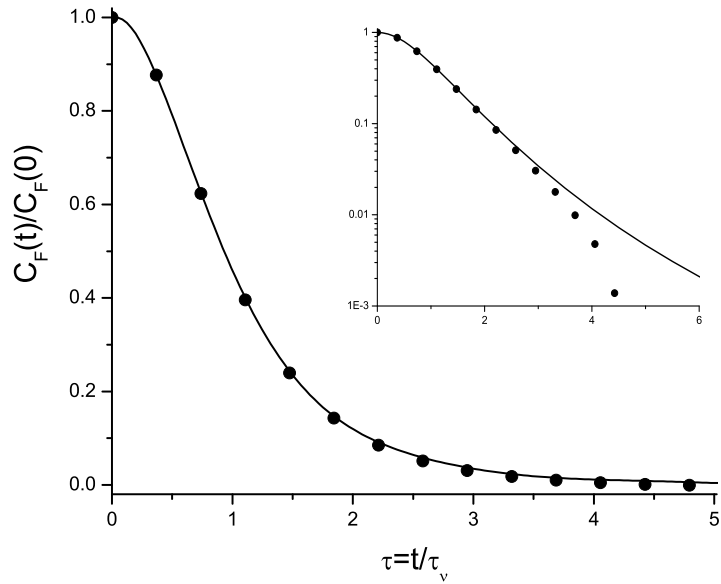}
\caption{Force autocorrelation function (FACF) as a function of 
$\tau=t/\tau_{\nu}$  (solid line) for packing fraction $0.3$ and $\nu=1152$, 
compared with MD simulations (circles). The insert shows the same except 
on a logarithmic scale, in order to make manifest that the asymptotic 
theory breaks down at large $\tau$.}
\end{figure}

\begin{figure}
\includegraphics[width=0.6\columnwidth]{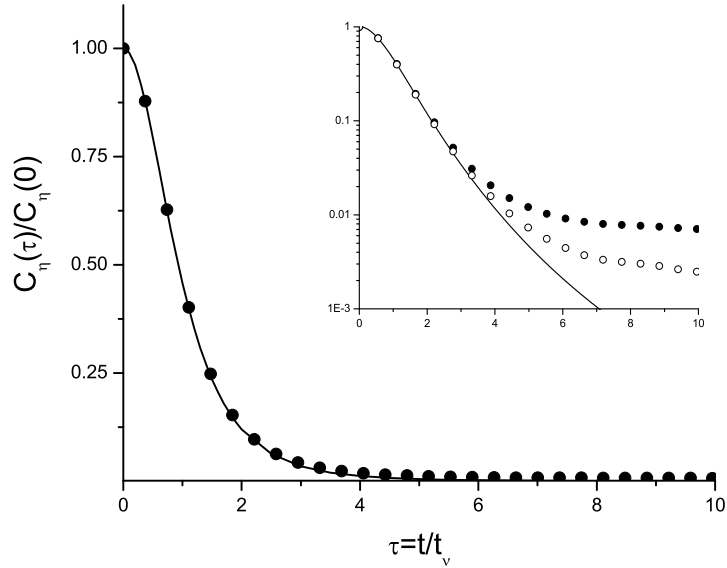}
\caption{Stress autocorrelation function (SACF) as a function 
of $\tau =t/\tau_{\nu}$ (solid line) for packing fraction $0.3$ 
and $\nu=1152$, compared to MD simulations (solid circles). The insert 
is a logarithmic plot of the same data, to show that the asymptotic theory 
breaks down for large $\tau$ values. Note that the MD data (open circles) 
for the collisional (vv) part of the SACF agree with the theory over a 
larger time interval than the full SACF, as is to be expected.}
\end{figure}

Consider first the FACF. Figure 3 shows that the agreement
between the short time crossover function and MD simulation is
excellent at very short times. The insert shows the same data on a
logarithmic scale to discover that deviations occur for $\tau \ge
3$. Figure 4 shows a similar representation for the SACF.
Also shown in the insert for this figure is the potential part of the correlation
function. It is seen that the agreement between the crossover
function and the potential part is again excellent up to about
$\tau = 4$.

The VACF shows a smoother crossover to the hard sphere behavior.
Figure 5 shows the good agreement for $\tau \leq 4$, and also
shows clearly the crossover to the hard sphere form, as   follows
from the long time expansion. The extrapolation of this hard
sphere form to $t=0$ gives a value different from $1$, due to the
crossover to soft sphere behavior near $t=0$. As $\nu $ increases
this short time domain goes to zero and the offset for the hard
sphere form goes to zero, as shown by the lower dashed curve. The
large$-\tau$ asymptote, $y = \tau - \half \sqrt{\pi}$, shows a
horizontal shift $\half \sqrt{\pi}$, as follows from the
asymptotic expansion \Ref{4.18}.
\begin{figure}  
\includegraphics[width=0.6\columnwidth]{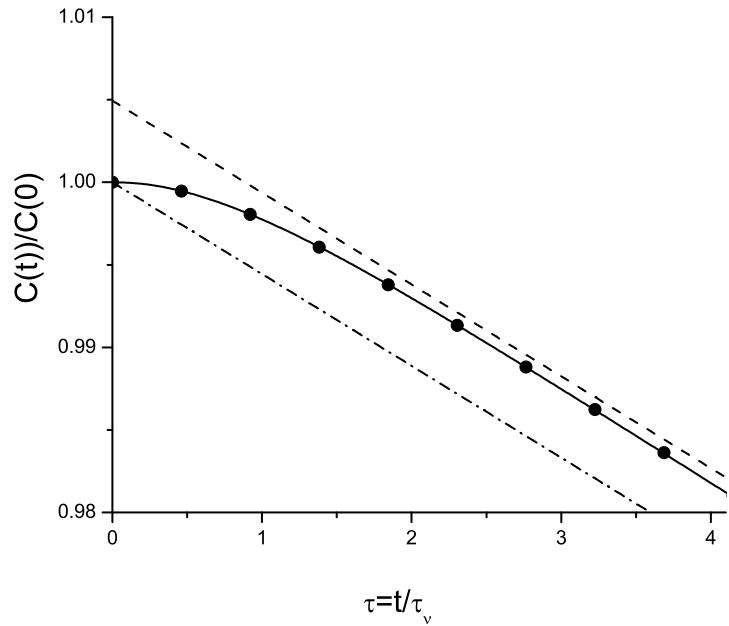}
\caption{Velocity autocorrelation function (VACF) as a function of $\tau =t/\tau_{\nu}$ 
for packing fraction $0.3$ and $\nu=1152$ (solid circles), 
compared to MD simulations (solid circles). Also shown is the same asymptote 
as in figure 2(b), parallel to the exact initial slope $\gamma_D$ and shifted 
by the constant "offset" in Eq.\Ref{4.18}.}
\end{figure}

\renewcommand{\theequation}{A.\arabic{equation}}
\setcounter{section}{6} \setcounter{equation}{0}
\section{Conclusions}

The exact short time dynamics has been calculated for the force,
stress, and velocity  autocorrelation functions for the special
case of a power law potential with large negative exponent. The
analysis applies for times much shorter than the mean free time
between pair  collisions $t_E$, but includes times less than and
greater than the time $\tau_\nu = t_\sigma/\nu$, which is the time
a particle needs to traverse the steep part of the potential. This
latter interval around $\tau_\nu$ is referred to here as the
crossover domain because the dynamics is characteristic of fluids
with smooth interactions at the shorter times $(t< \tau_\nu)$, but
becomes more representative of a hard sphere fluid at longer times
$( \tau_\nu<< t << t_E)$. The results can be all expressed in a
single universal scaling functions, $\sS(\nu t/t_\sigma)$, which
is the autocorrelation function of the pair force,
$\avo{F_{12,x}(0) F_{12,x} (t)}$ at times $t<< t_E$. It is
independent of the dimensionality of the system.

Thus, identification of the crossover dynamics allows several
conceptual and quantitative questions to be addressed regarding
the relationship of fluids with steeply repulsive interactions to
those of the hard sphere fluid. To pose such questions it is
important to have detailed results for the hard sphere fluid.
There are complications due to the singular nature of the
interactions. For example, the correlation functions associated
with the Green-Kubo expressions for transport coefficients involve
fluxes with forces that are ill-defined for hard spheres.
Similarly, the usual Hamiltonian dynamics for these correlation
functions is no longer applicable for hard spheres.

The first part of our analysis here was to show how these problems
could be handled, with well defined expressions for the hard
sphere fluid. First, the Helfand representation for transport
coefficients was recalled as an equivalent alternative to the
Green-Kubo forms. It was shown in Appendix A how this
representation could be used, together with an appropriate
pseudo-Liouville dynamics to arrive at the corresponding
Green-Kubo expressions for the hard sphere fluid. Interestingly,
these differ in form from those for smooth interactions by
constant contributions due to instantaneous collisional transfer
contributions from initial configurations of particles at contact.
This implies that there appears in the hard sphere limit $( \nu
\to \infty)$ a delta function singularity in the correlation
functions starting with smooth interactions. The exact short time
behavior of the velocity autocorrelation function and
kinetic-kinetic parts of the Green-Kubo time correlations
functions, are described by the scaling function $\sF( \nu
t/t_\sigma)$, which satisfies $d^2 \sF (\tau)/ d\tau^2 = - \sS
(\tau)$. In the hard sphere limit these functions have cusp
singularities, with exact short time expansions, depending
linearly on the magnitude of $t$, while the corresponding
functions for the smooth interactions are analytic in time.

A primary result reported here is the identification of a single
crossover function $\sS(\tau)$ that describes the short time
dynamics of all of these time correlation functions for smooth
interactions. The above anomalous differences between smooth and
hard interactions can be understood qualitatively and
quantitatively from the properties of this function. For the
stress autocorrelation function it yields a delta function
singularity showing the consistency between the different
Green-Kubo forms for smooth and hard interactions. It also shows
the non-uniform nature of the short time limit with respect to the
steepness of the potential, explaining the differences between the
analytic and non-analytic forms of the short time expansions for
smooth and hard interactions, respectively.

\renewcommand{\theequation}{A.\arabic{equation}}
\setcounter{section}{7} \setcounter{equation}{0}
\section{Acknowledgements}

The research of JWD was supported by Department of Energy grant DE-FG02-02ER54677.

\renewcommand{\theequation}{A.\arabic{equation}}
\setcounter{equation}{0}

\section*{Appendix A: Pseudo-Liouville and
Binary Collision Operators}

The time evolution of any phase function $A(t)$, evolving under
hard sphere dynamics with diameter $\sigma$ can be described for
positive times $(t>0)$ by the forward generator $A(t) = e^{tL_+}
A(0) $ with the pseudo-Liouville operator $L_{+}$ \cite{ErnstHS}.

The equilibrium time correlation functions necessarily satisfies
the relation $\avo{A(0) B(t)} = \avo{B(0) A(-t)}$ because of
stationarity. This requires in the hard sphere case a {\it
backward} generator $e^{-tL_-} (t>0)$, where
\ba \Label{c2}
&L_\pm = L_0 \pm \sum_{i<j} T_\pm (ij)& \nn  &L_0 = \sum_i \bv_i \cdot
{\bf \nabla_i}&
\nn &T_\pm (ij) = \sigma^{d-1} \int^{(\mp)}\!\!
 d \hbsig \,|\bg_{ij} \cdot \hbsig|
\,\delta( \br_{ij}-\bsig) (b_\sigma -1)&.
\ea
The superscript on the $\hbsig-$integral denotes the constraints
($\mp \bg_{ij}\cdot \hbsig >0)$. The $b_\sigma-$operator is a
substitution operator, executing the collision laws, i.e.
\ba \Label{c3}
&b_\sigma \bv_i = \bv_i^\prime &= \bv_i - (\bg_{ij}\cdot
\hbsig)\hbsig
\nn &b_\sigma \bv_j = \bv_j^\prime &= \bv_j + (\bg_{ij}\cdot
\hbsig)\hbsig.
\ea
The stationarity of the time correlation function under hard
sphere dynamics for positive $t$ is expressed as,
\be \Label{c4} \avo{Be^{tL_+} A} \equiv \int d
\Gamma \rho_o Be^{tL_+}A = \int d\Gamma \rho_o Ae^{- tL_-}B \equiv
\avo{Ae^{-tL_-}B},
\ee
where the operators $L_\pm$ and $e^{\pm t L_\pm}$ inside the
thermal averages $\avo{ \cdots }$  are always preceded by an
$N-$particle equilibrium distribution function $\rho_o$. It
includes the overlap function $W(N)$, which vanishes for
overlapping configurations (with at least one pair distance
$r_{ij} < \sigma$), and $W(N)=1$ for non-overlapping
configurations. For the derivation of these and other properties
we refer to the original
literature \cite{ErnstHS}. \\
The only additional properties needed in this article are the
conservation laws for the summational invariants $a_i =\{ 1,\bv_i,
v_i^2 \}$, where
\be \Label{c5}
T_\pm (ij) (a_i +a_j) = 0.
\ee

An important application of \Ref{c4} in the present context is the
interpretation of the Einstein-Helfand formulas \Ref{2.9} with the
time correlation function $ \avo{(M(t)-M(0))^2}$ for hard sphere
fluids. There is a caveat here. The phase function with $A(t)B(t)$
and $t>0$ can only be expressed in hard sphere generators as
$e^{tL_+} A(0)B(0)$. It is {\it not} equal to $(e^{tL_+}
A(0)(e^{tL_+}B(0))$, because $e^{tL_+}$ is not a substitution
operator, like the streaming operator $e^{tL}$ for smooth
interactions. One shows stationarity of an equal-time average as
$\avo{A(t)B(t)} = \avo{e^{tL_+}AB} = \avo{AB}$, where we have used
\Ref{c4} and the relation \Ref{c4} in the form $e^{-tL_-} 1=1$.

With this in mind the Helfand formula \Ref{2.9}, say for the shear
viscosity, can be expressed in hard sphere generators as,
\ba \Label{c6}
\eta_V(t) &=&\frac{\beta}{V} \left(\frac{d}{dt}\right) \avo{M^2(0)
- M(0)M(t)}
\nn &=& - \frac{\beta}{V} \left(\frac{d}{dt}\right) \avo{M
e^{tL_+} M} =- \frac{\beta}{V} \avo{M e^{tL_+}L_+ M}.
\ea
Application of \Ref{c4} and use of the identity,
\be
e^{-tL_-} = 1-\int_0^t ds e^{-sL_-}L_-,
\ee
allows us to transform the Helfand formula for hard spheres
into,
\be \Label{c8}
\eta_V(t) = - \frac{\beta}{V} \avo{M L_+ M} + \frac{\beta}{V}
\int_0^t ds \avo{(L_- M) e^{s L_+} L_+ M}.
\ee
This relation is used in \Ref{3.2} of the main text.

\end{document}